# Protrusion fluctuations direct cell motion

*David Caballero [1,2], Raphaël Voituriez [3,4], Daniel Riveline [1,2,*]*

[1] Laboratory of Cell Physics, ISIS/IGBMC, Université de Strasbourg and CNRS (UMR 7006), 8 allée Gaspard Monge, 67083 Strasbourg, France

[2] Development and Stem Cells Program, IGBMC, CNRS (UMR 7104), INSERM (U964), Université de Strasbourg, 1 rue Laurent Fries, BP10142, 67400 Illkirch, France

[3] Laboratoire de Physique Théorique de la Matière Condensée, CNRS UMR 7600, Université Pierre et Marie Curie, 4 Place Jussieu, 75005 Paris, France

[4] Laboratoire Jean Perrin, CNRS FRE 3231, Université Pierre et Marie Curie, 4 Place Jussieu, 75005 Paris, France

[*]Corresponding author: Daniel Riveline; e-mail: riveline@unistra.fr; Laboratory of Cell Physics, ISIS/IGBMC, Université de Strasbourg and CNRS (UMR 7006), 8 allée Gaspard Monge, 67083 Strasbourg, France. TEL: +33 (0) 3 68 85 51 64; FAX: +33 (0) 3 68 85 52 32.

Condensed running title: Protrusion fluctuations direct cells

**Abstract**
Many physiological phenomena involve directional cell migration. It is usually attributed to chemical gradients *in vivo*. Recently, other cues have been shown to guide cells *in vitro*, including stiffness/adhesion gradients or micro-patterned adhesive motifs. However, the cellular mechanism leading to these biased migrations remains unknown, and, often, even the direction of motion is unpredictable. In this study, we show the key role of fluctuating protrusions on ratchet-like structures in driving NIH3T3 cell migration. We identified the concept of efficient protrusion and an associated direction index. Our analysis of the protrusion statistics facilitated the quantitative prediction of cell trajectories in all investigated conditions. We varied the external cues by changing the adhesive patterns. We also modified the internal cues using drug treatments, which modified the protrusion activity. Stochasticity affects the short- and long-term steps. We developed a theoretical model showing that an asymmetry in the protrusion fluctuations is sufficient for predicting all measures associated with the long-term motion, which can be described as a biased persistent random walk.

**Introduction**
Many physiological processes, such as tissue development or immune response (1, 2), as well as some pathological phenomena, such as tumor invasion or cancer metastasis (1-4), involve cell migration. Various studies have reported that this phenomenon is mainly a result of the chemical gradients that lead to cell polarization and the regulation of signaling networks (5, 6), although the gradients were not reported systematically. Other cues were also shown to direct cell (fibroblast and endothelial) motion (7-11). For example, human endothelial cells migrate directionally towards regions of higher concentrations on surfaces with gradients of adhesive proteins. Similarly, on gradients of substrate rigidity, fibroblasts move towards regions of higher rigidity (7, 12). However, in general, cells do not move along directions that are set by these simple situations, and this situation prevents the quantitative prediction of cell motion.

Locally, many cells probe their environments through extensions called protrusions: actin gels grow from the cell edges, and cells extend their borders through filopodia and lamellipodia. Protrusions grow and shrink stochastically around the cell on timescales of minutes and lengths of micrometers. When protrusions are eventually stabilized, adhesion is triggered locally, and a local force is applied by the cell. If the cell is polarized, an imbalance between the protrusions at the cell ends may lead to a directed motion. The onset of cell polarization and directed motion therefore seems to involve fluctuations in protrusions. In fact, filopodia dynamics was shown to play a key role in the turning of nerve growth cone to face a chemical signal to connect to a specific partner cell (13-15). However, evidence that an asymmetry in protrusion activity is a predictor for the long-term cell migration direction is lacking so far.

More generally, fluctuations have been shown to play an essential role in many biological systems, such as molecular motors (16). This idea was pioneered by Richard Feynman in his ratchet and pawl chapter (17), where he showed that the non-directional motion driven by fluctuations is rectified by breaking temporal and spatial symmetry. Inspired by this framework, we aim to understand how the fluctuations of protrusions regulate directional cell motion. In particular, we examined how NIH3T3 cells behave in environments where only protrusion activity triggers cell motility without other regulatory mechanisms, such as chemoattractants. For that purpose, we plated NIH3T3 cells on a series of adhesive patches that had asymmetric triangular shapes (Fig. S1(a)). These adhesive patches were separated by non-adherent gaps. This set-up provided an asymmetric guide for the growth and dynamics of cell protrusions, mainly filopodia, towards the neighboring triangles.

We quantified stochasticity by measuring the frequencies of the extension and adhesion of the protrusions. We found that the cells extended protrusions more frequently from the broad end of the triangular patch than from its pointed end, whereas the filopodia extending from the pointed end were more stable than those from the broad end. As a result, cell motion was possible in either direction; however, on average, the cells migrated mostly towards the direction defined by the pointed end in both short- (10 h) and long-term experiments (days) - a relevant timescale for development or physiological processes. Furthermore, by regulating the cytoskeleton dynamics by inhibiting the Rho and Rac pathways, we altered the nature of the protrusion fluctuations and modified the motion of the cells on the same ratchets. In all cases, we could define and measure the frequencies of probing and adhering.

We developed a simple mesoscopic model of a persistent random walk, using the experimentally measured biased probabilities of protruding and adhering as inputs. We obtained excellent quantitative agreements for the direction, long-term ratchet efficiency and persistence in motion. These results demonstrate that the asymmetries in the frequency and stabilization time of protrusions are key physical factors in setting cell direction.

**Material and Methods**

*Micropattern fabrication:* Microcontact printing was used for fibronectin micropatterning. Poly(dimethylsiloxane) (PDMS) (Sylgard 184, Dow Corning, USA) stamps (pre-polymer:cross-linker, 10:1 (w/w)) were replicated from a SU-8 mold (MicroChem Corp.) fabricated by standard UV photolithography (MJB3 mask aligner; SUSS MicroTec, Germany). The stamps were rendered hydrophilic by $O_2$ plasma treatment. Then, they were inked for 60 min with a 10 µg ml$^{-1}$ rhodamine-labelled fibronectin solution (Cytoskeleton Inc., USA) (18), dried and placed in contact with a #1 glass coverslip (Marienfeld GmbH & Co., Germany) for 5 min, which had been previously functionalized with 3-(mercapto)propyltrimethoxysilane (Fluorochem, UK) by vapor phase for 1 h and cured for 4 h at 65°C (see Fig. S1) (19). After releasing the stamp, we cleaned the samples by immersion in PBS (pH 7.4), Milli-Q water and 10 mM Hepes (pH 7.4) solutions. Non-functionalized regions were blocked using 0.1 mg ml$^{-1}$ solution of poly-L-lysine-g-poly(ethylene glycol) (PLL-g-PEG) (SurfaceSolutions GmbH, Switzerland) in 10 mM Hepes (pH 7.4) at room temperature for 30 min. Finally, the samples were rinsed with PBS twice and stored in PBS at 4°C before cell deposition.

*Cell culture:* Mouse NIH3T3 fibroblasts (ATCC, USA) were grown in high-glucose Dulbecco's Modified Eagle's Medium (Invitrogen, France) supplemented with 1% Pen Strept antibiotics (Invitrogen) and 10% bovine calf serum (BCS) (Sigma-Aldrich, France) at 37°C and 5% $CO_2$. The cells were trypsinized (0.25% Trypsin-EDTA) (Invitrogen, France), centrifuged and deposited on the microcontact printed substrate at a low density (1-2x10$^4$ cells·cm$^{-2}$) to reduce cell-cell interactions. After 20 min, the medium was replaced with fresh medium to remove non-adherent cells. Finally, for the experiments, we used an L-15 medium with a small amount of serum (1% BCS) to reduce the deposition of ECM proteins around the micropattern.

*Cytoskeleton drugs:* Cells were incubated with 80 nM of C3 transferase (Cytoskeleton Inc., USA) and 100 µM of NSC23766 (Calbiochem, France) (20, 21).

*Optical microscopy:* Short-term cell images were acquired with an inverted microscope (Olympus, Japan) (40X 0.65 N.A. phase-contrast air objective, 1 image/30 s). The microscope was equipped with a CCD camera (Hamamatsu, Japan), an Hg lamp (FluoArc, Zeiss, Japan) for epifluorescence experiments, 2 shutters (Uniblitz, USA) and a red filter (ThorLabs Inc., USA) to prevent phototoxicity. For long-term experiments, a 4X 0.25 N.A. phase-contrast objective was used (1 image/5 min). A thin layer of mineral oil (Sigma-Aldrich, France) was used to cover the medium to prevent evaporation.

*Main parameters of the protrusion activity:*

- The frequency of probing $\nu$ is defined as the number of filopodia that reach an adhesive fibronectin area A per unit of time (see Fig. S2).

- The stabilization time $\tau$ is defined as the dwelling time of a protrusion (filopodia) on an adhesive fibronectin area (see Fig. S2).

- A protrusion is considered efficient if it leads to force transmission.

- The quantity z denotes the number of efficient protrusions that are generated per unit of time. Assuming that a protrusion becomes efficient with rate $\beta$ while in contact with an adhesive area, it is shown below that $z \approx \beta \nu \tau$.

Note that the quantities $\nu$, $\tau$ and z depend on the direction of motion s specified below. Filopodia protrusions were observed and measured only at the initial stage prior to the first-step motion. The quantities $\nu$ and $\tau$ were both manually measured from time-lapse movies using ImageJ software, whereas z cannot be accessed directly experimentally. The adhesive area $A_\pm$ is defined as the intersection between the protrusion exploring area and the fibronectin adhesive motifs and depends on the direction, +/−. NIH3T3 cells were allowed to completely spread on the FN motif prior to the start of the analysis and until the cell was about to migrate. Finally, $d_p$ is defined as the average protrusion (filopodia) length. The acquisition rate was 1 image/30 s. A 40X 0.65 N.A. phase-contrast air objective (Olympus) was used.

*Long-term biophysical parameters*: Cell trajectories were manually tracked (Manual Tracking Plug-in, ImageJ) at 5 min intervals for 48 h. Points were connected to generate a set of migration paths, which were used to calculate the averaged values of cell persistence length $L_p$, persistence time $T_p$, velocity v and pausing time $T_{pa}$. $L_p$ and $T_p$ are defined as the length and time during which the cell moves straight without stopping, respectively, and v is the speed of this motion. Note that each cell could have as many $L_p$, $T_p$ and v values as straight paths for its total trajectory. The resulting parameters were obtained by averaging out all the individual values for all the trajectories and cells. For the velocity measurements, the pausing times $T_{pa}$ of the cells were excluded. $T_{pa}$ is the time during which a cell does not move in a motif. We considered cells to be pausing if they did not move for t>30 min. The number of turns per unit time $Nt^{-1}$ was recorded as the number of turns that the cells had performed during their full motion. To classify motions into +, − or null, the cell positions were compared between the start and the end of movies. Null (no net motion) corresponded to cells returning to the original pattern location, although the cells had been fluctuating with +/− motions. Data are provided as the mean ± SEM. Statistical analysis was performed using Student's t-test, and significance was accepted at P<0.05.

**Results and Discussion**

We designed an assay in which we plated NIH3T3 fibroblasts on a series of adhesive micropatterned areas (fluorescent fibronectin, Cytoskeleton, 10 µg ml$^{-1}$) (see Fig. 1(a)-(b)) (18, 19, 22). We selected this fibronectin concentration to obtain large cell velocities while allowing the cells to crawl on surfaces. This allows cells to protrude while being completely spread on the FN motifs and to migrate to multiple motifs during the time of the experiments (see Fig. S2). This velocity is close to the maximum velocity in the dumbbell motility curve (23) (see Supplementary Material and Methods and Fig. S3 in the Supporting Material). The area of each patch, 1590 μm², corresponds to the mean NIH3T3 cell area on non-patterned surfaces (see Fig. S1(b)). The structures have asymmetric triangular shapes; they are separated by non-adherent gaps of 20.5 μm, corresponding to the average protrusion (filopodia) length $d_p$ (see Fig. S1, S4 and Supplemental Methods). The non-patterned regions were passivated by the repellent PLL-g-PEG (0.1 mg ml$^{-1}$), forcing the cells to extend protrusions to probe the adhesive regions at a distance $d_p$ (Fig. 1(c) and Supplementary Methods). This gap feature is in contrast to the situation of cells on uniform 2D surfaces and on micropatterned surfaces of connected adhesive patches (24); these cells show continuous protrusion-retraction cycles. This set-up provided asymmetric guidance for the growth and dynamics of the cell protrusions towards the neighboring triangles.

To move to neighboring adhesive patches, cells must extend filopodia that are longer than the gap. This implies that the direction of cell motion may be determined by the asymmetry of protrusions. We therefore observed filopodia with an inverted microscope (40X 0.65 N.A. phase-contrast air objective, 37°C, 1% bovine calf serum (BCS) in L-15 medium (Invitrogen)) (see Fig. 1(c), Movie S1). Our first measurements indicated that the cells extended filopodia protrusions with similar lengths $d_p$ in both directions, defined as +/− directions (see Fig. 1(c), Fig. S4). We assumed that the number of efficient protrusions (defined as protrusions that eventually lead to the transmission of a force, see Material and Methods) that are generated per unit of time z is proportional to the intersection of the explored area and the fibronectin pattern (effective adhesive area) (see Fig. 1(c), dashed lines). This choice is motivated by the distribution of filopodia protrusions distances: multiple protrusions of various lengths explore the neighboring adhesive area at the same time. To bias the activity of efficient protrusions, we designed the ratchet so that the effective adhesion areas differed on the two sides ($A_+ \neq A_-$), leading to an asymmetry in the number of efficient protrusions generated per unit of time, $z_+ \neq z_-$ (see Fig. 1(c)). We then measured the frequency of probing of the filopodia that contacted the adhesive zones for each of the two sides ($\nu_+$ and $\nu_-$) (see Fig. S2). We found that they were more numerous on the − side (see Fig. 2(a)i), but that they lasted longer on the + side (dwell time on adhesive patches are denoted $\tau_+$ and $\tau_-$). This suggests that $\nu_\pm$ and $\tau_\pm$ are two competing factors that control the activity of efficient protrusions. To evaluate the normalized rate of filopodia extensions, we measured the frequency of probing per unit of effective adhesive area ($A_-, A_+$) (data not shown). We observed that a higher rate of filopodia was obtained on the − side. This is a consequence of constraining the cell shape in a triangular geometry (25); the cells oriented their stress fibers (and focal contacts) (see Fig. 1(a)) on the +/− sides. As shown in Fig. 1(a), the number of stress fibers and paxillin is larger on the wide edge, which is consistent with a higher rate of filopodia.

Assuming that a protrusion can be activated (*i.e.*, generate a force) at a constant rate β (assumed constant for the sake of simplicity) when it lies on an adhesive area $A_\pm$, the number of efficient protrusions generated per unit of time is given by $z_i = s_i \nu_i \approx \beta \nu_i \tau_i$ and depends on the side i=+,− ($s_i$ is the probability that a protrusion is activated on side i). For simplicity, we assume that all the filopodia apply the same force. This is a plausible assumption based on the well-established fact that focal contacts, which nucleate and grow at this timescale, mediate the traction force (26). *Note that we are extending the classical definition of filopodia as searching organelles to efficient protrusions, a more integrated structure which integrate probing and adhering.* We then introduce the following parameter, the 'direction index' ($I_{dir}$):

$$I_{dir} = (z_+ - z_-)/(z_+ + z_-) \quad (1)$$

that describes the asymmetry of efficient protrusions and depends only on the parameters $v_i$ and $\tau_i$, which are accessible experimentally. We show in the Supplemental Methods that $I_{dir}=p_+ - p_-$, where $p_+$ is the probability that an efficient protrusion is on side + rather than side −. Our central hypothesis is that the average direction of an elementary step from one pattern to a neighboring one is dictated by the asymmetry in the number of efficient protrusions, which is directly quantified by $I_{dir}$.

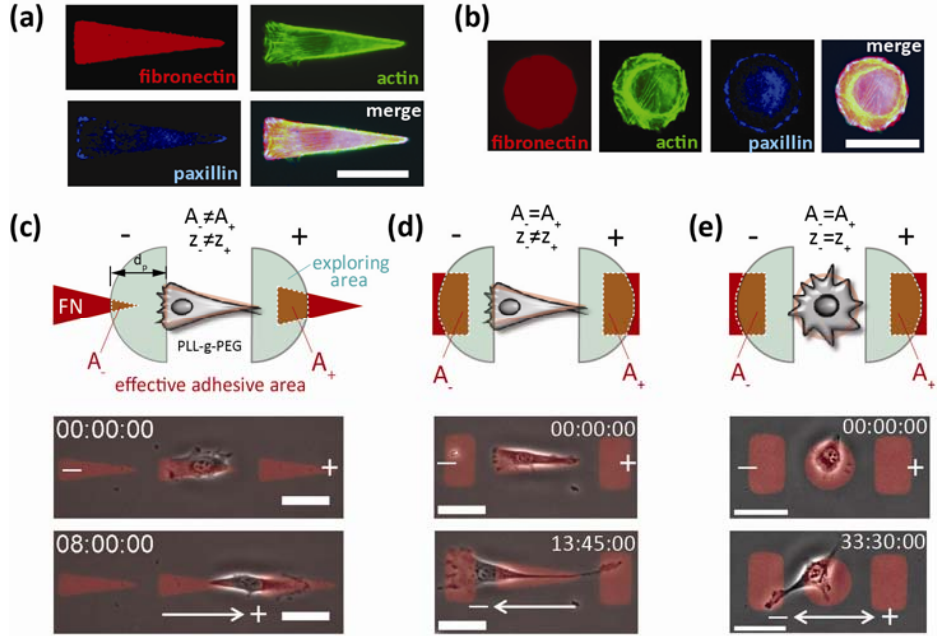

**Fig. 1.** Cells on our patterns. **(a, b)** NIH3T3 cells on triangular and circular fibronectin patches, respectively, with staining for actin and paxillin. *(c, d, e) (Upper)* Scheme for probing the activity for all the studied configurations showing the relevant parameters. $A_-, A_+$ and $z_-, z_+$ represent the efficient adhesion area (white dashed lines) and filopodia protrusion activity (frequency of probing and stabilization time of protrusions on $A_\pm$), respectively, in the + and − directions, as indicated. The cell exploration area is shown in green. The parameter $d_p$ represents the average protrusion length. *(Lower)* Typical first-step motion for the different studied configurations. Time in hh:mm:ss. Scale bars: 50 µm.

We found that the NIH3T3 cells migrated in the − direction if $I_{dir}<0$, in the + direction if $I_{dir}>0$ or in any direction if $I_{dir}\approx0$. The limits of this approximation are set by the corresponding distribution of the data (error bars) within the + and − regions. The sign of $I_{dir}$ is therefore directly correlated to the direction of motion: this key parameter determines the direction of cell migration (see Fig. 2(b)-2(d)). The ratchet configuration yielded $I_{dir}=0.33$ (see Fig. 2(b)), showing that filopodia extensions are more efficient in the + direction (see Movie S1), which is in agreement with the actual direction of migration. Surprisingly, despite this direction of motion, the NIH3T3 cells are initially polarized towards the − direction, as confirmed by the focal contacts distributions and their mean asymmetric areas (Fig. 2(c)); the position of the centrosome with respect to the nucleus also suggested this polarity tendency (see Fig. S5 (27)). However, the accessible adhesive area is larger in the + direction; thus, eventually, the efficient protrusions were more numerous at the + edge (see Fig. 1(c)). Hence, although the NIH3T3 cell morphology initially followed the geometry of the triangular motif, these cells reversed their polarity when they migrated in the + direction (see Fig. 1(c) and Movie S1).

To further test these ideas, we modified the geometry of the motifs while maintaining the gap distance. We first used a pattern where a triangle is surrounded by symmetric rectangles, and hence, the cells have equal available adhesive area ($A_+=A_-$) but an asymmetric protruding distribution ($v_->v_+$) (Fig. 1(d), Fig. 2(a)ii, and Movie S2). As expected, the frequency of efficient protrusions is larger towards the − direction than towards the + direction, $z_->z_+$, (see Fig. 2(a)ii). Indeed, the protrusions are stabilized for longer times on the − side than on the + side, $\tau_->\tau_+$. This results in a negative value of $I_{dir}$ (-0.57) and predicts that cells migrate towards the − direction, in agreement with the experimental observations (see Fig. 2(b)-2(d)).

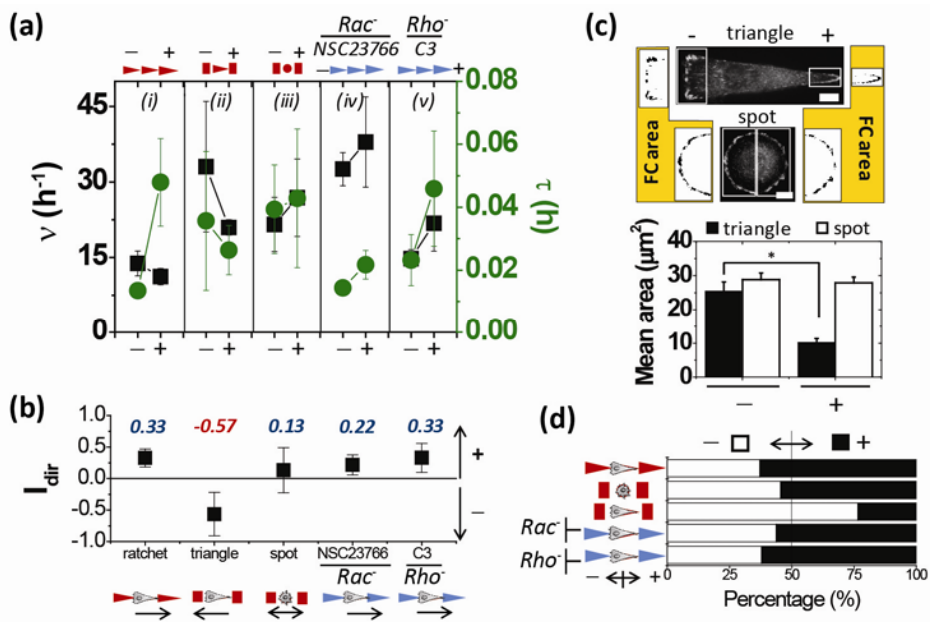

**Fig. 2.** Protrusion (filopodia) activity and first-step cell motion. **(a)** Measurement of the frequency of probing $v$ and the stabilization time $\tau$ of filopodia in the + and − sides. The data show an asymmetry in the filopodia dynamics for cases i, ii, iv and v. **(b)** The direction index $I_{dir}$ was calculated using the measured $v$ and $\tau$ and Eq. 1. The $I_{dir}$ parameter shows a correlation between the sign and the migration direction. (Data set: $n_{ratchet}$=4; $n_{triangle}$=4; $n_{spot}$=5; $n_{C3}$=3; and $n_{NSC23766}$=4). **(c)** *(Upper)* Labeling of the focal contacts (paxillin) and mean area quantification for the evaluation of cell polarity. Scale bar: 10 µm. *(Lower)* A bias in the focal area was obtained for the triangle compared to the spot, confirming the polarization of the cells (Data set: $n_{spot}$=25; $n_{triangle}$=23). **(d)** First-step migration assay. WT, Rac⁻ and Rho⁻ cells in the ratchet configuration show a first-step migration mainly towards the + direction. When replacing the neighbor motifs by symmetric rectangles for the WT, the protrusion dynamics are altered, as shown in (a)ii. In this case, the direction of migration is switched towards -. No net bias is observed for the control case. (Data set: (n;N)$_{ratchet}$=43;5. (n;N)$_{spot}$=46;2. (n;N)$_{triangle}$=39;2. (n;N)$_{NSC23766}$=96;3. (n;N)$_{C3}$=29;4). The data are presented as the mean ± SEM. *P<0.001 (Student's t-test).

Replacing the triangle by a circular patch (spot) made the pattern +,− symmetric ($A_+=A_-$) (see

Fig. 1(e)). In this situation, $v_- \approx v_+$ and $\tau_- \approx \tau_+$ within experimental error; thus $z_+ \approx z_-$ (see Fig. 2(a)*iii* and Movie S3). In this case, $I_{dir} \approx 0$ within the error (Fig. 2(b)). As above, $I_{dir}$ correlated with the obtained average motion: here, + and − were equally probable. $I_{dir}$ therefore sets the direction of motion in all conditions.

We next perturbed the fluctuations of the protrusions in the ratchet by using inhibitors of the Rac and Rho pathways (hereafter written as Rac⁻ and Rho⁻ conditions, respectively), which are known to control the dynamics of the cytoskeleton (28) (see Fig. 2(a)*iv-v*). We used 80 nM C3 Rho-inhibitor (Cytoskeleton) (20) and 100 μM NSC23766 Rac-inhibitor (Calbiochem) (21). Accordingly, $v$ and $\tau$ were altered compared to the values for the wildtype (WT) cells. Cells probed the + edge more often in both cases ($v_- < v_+$). Similarly, we found $\tau_- < \tau_+$, which leads to positive $I_{dir}$ values for both cases (0.22 and 0.33 for Rac⁻ and Rho⁻, respectively) (see Fig. 2(b)). This value predicts that NIH3T3 cells migrate towards the + direction, in agreement with the experimental observations (Fig. 2(d)). Surprisingly, we did not observe significant differences in the directionality of the single-step motion of the cells after the Rac⁻ and Rho⁻ treatment (see Fig. 2(d)) although $v$ and $\tau$ were altered. This suggests that the FN patterns constrain the cells and impose the behavior for the first step. After this step, the cell shape is less constrained and does not conform to the shape of the patterns, and the inhibitions start to show their long-term effects. Altogether, the first-step motion of the cells confirmed the prediction that $I_{dir}$ dictates the direction of motion. When cells are polarized, they probe their front environment more frequently. However, the probability of finding a stable attachment site is also critical in setting the cell direction. The important feature is the asymmetry in the number of efficient protrusions described in $z_+$ and $z_-$ and therefore in $I_{dir}$. In the case of the ratchet pattern, attachments form more easily on the backward (+) side, and this difference is sufficient to reverse the direction of motility.

We next studied the long-term motion for 48 h with a low-magnification objective (4X 0.25 N.A. phase-contrast). On a line of multiple spots, wild-type cells are able to hop from one patch to the next, as shown in Fig. 3(a) (see also Movie S4). While sitting on a patch, the cells are not polarized (Fig. 1(b)). As expected, we obtained no bias in this condition (Fig. 4(a)). We then used triangular patches; a variety of behaviors were observed, ranging from fluctuating cells (Fig. 3(b)*i*), cells that moved directionally towards either the left or right along the ratchet (Fig. 3(b)*ii*) and cells that did not move (approximately 5% of the cells). In some cases, the cells were also initially fluctuating in the +/− directions; then, they exhibited directed motion (Fig. 3(b)*ii* and Movie S5). The average motion is significantly biased in the + direction, as demonstrated by the proportion of cells moving in the + direction over those moving in the − direction (+/−). In this case, +/− was 2.5-fold the value obtained for non-polarized cells on circular patches (see Fig. 4(a)).

If the protrusion activities play a key role, then perturbing these protrusions should affect the long-term motion. To test this hypothesis, we changed the protrusion dynamics by probing the Rho⁻ and Rac⁻ conditions (28, 29). The cytoskeleton was modified (see Fig. S6). Surprisingly, the +/− proportion was not significantly altered (Fig. 4(a)). However, the characteristics of the trajectories changed. Rho⁻ cells were less polarized: the cells were less persistent (lower $L_p$) and stopped more often (see Fig. 4(b) and Fig. S7). The cells showed a reduced speed most likely due to a decrease in the number of stress fibers and the associated applied forces (see Fig. S6). However, the asymmetry of the ratchet yielded a stronger asymmetry in the efficiency of protrusions (Fig. 2(a)v): the probability to go + was larger than the probability to go −, eventually yielding a +/– ratio similar to that of the untreated case. In contrast, Rac⁻ enhanced the cell polarity, and the $L_p$ values were therefore larger in both directions. The polarity was strong enough so that upon polarization in the – direction, a cell maintained this direction for longer distances than in the untreated case, eventually yielding a similar +/– ratio (Fig. 4(a)). Surprisingly, the cells with decreased Rac activity showed higher speeds than the WT speed. This might be a consequence of a reduction of peripheral lamellipodia (30). This prevents cells from exploring sides and reduces the chances to change directions, thereby increasing the speed and directionality of cells in our setup. Additionally, the speed values can vary depending on the level of Rac activity and/or the cell morphology. Depending on this level, the cells that show the highest directionality and speed and have a spindle-shaped phenotype with a stable lamella in the direction of migration, in agreement with our results (see Fig. S6) (30). Note also that Rac⁻ can lead to increased Rho activity (31): actomyosin contractility is increased, with an associated increase in the applied force and cell speed.

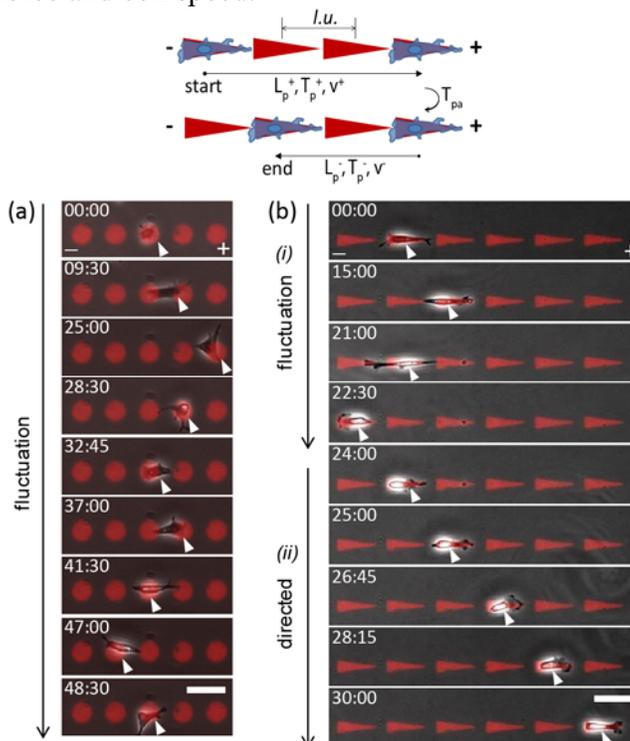

**Fig. 3.** Long-term NIH3T3 cell motion. *(Top)* Scheme describing the long-term biophysical parameters of persistence length $L_p$, persistence time $T_p$, speed v and pausing time $T_{pa}$ for the +/– directions. The lattice units are represented as ´l.u.´. **(a)** Time-lapse sequence for 48 h of a NIH3T3 fibroblast fluctuating on circular fibronectin patches. **(b)** Time-lapse sequence for 30 h of a NIH3T3 fibroblast *(i)* fluctuating on the ratchet and *(ii)* migrating directionally towards the + end. Scale bars: 100 μm. Time in hh:mm.

Note that our approach does not address how the Rho pathways are modified or antagonized but does address how a rectification leads to similar results even when two antagonistic pathways are challenged (32). This shows that opposite effects can lead to the same rectification. Rac⁻ cells probe more and adhere less than wildtype cells; they also have a larger $L_p$. Nevertheless, the rectification is quantitatively the same. Rho⁻ shows similar behavior. Beyond the molecular details of the pathways, the cellular events, such as polarity, protrusions, adhesions, can be integrated to lead to the same quantitative read-out because they compensate for each other. Altogether, this demonstrates that a perturbation of the protrusion dynamics has a direct effect on the parameters associated with long-term cell motion.

Other features, such as the ratios $L_p^+/L_p^-$ and $T_p^+/T_p^-$, the switching times ($Nt^{-1}$) and the pausing time ($T_{pa}$), also supported this framework (see Fig. 4(b)-4(c) and Fig. S7). $Nt^{-1}$ decreased for the triangular patches compared to that for the spot configuration (see Fig. S7), suggesting that the cell polarity is stabilized in this configuration, and as a consequence, the cell motions are more directional. This result is also in agreement with the focal contacts distribution (see Fig. 1(h)). Upon Rho/Rac inhibition, $Nt^{-1}$ was further reduced, suggesting a more stable polarity. This parameter is distinct from $L_p$ because, for example, the directionality ($L_p$) only increased in the Rac$^-$ condition. In addition, the pausing time $T_{pa}$ is related to the time needed to establish polarity through cytoskeleton reorganization and to the ability of cells to apply traction forces via stress fibers. For Rho$^-$, the high values of $T_{pa}$ could suggest lower traction forces (decrease in stress fibers) (see Fig. S6), consistent with a lower v. Rac$^-$ shows a $T_{pa}$ similar to that of the WT with a decrease in $Nt^{-1}$. This suggests a similar stability in polarity with a shorter period for establishing polarity, causing larger $L_p$ values, which is in agreement with the experimental data.

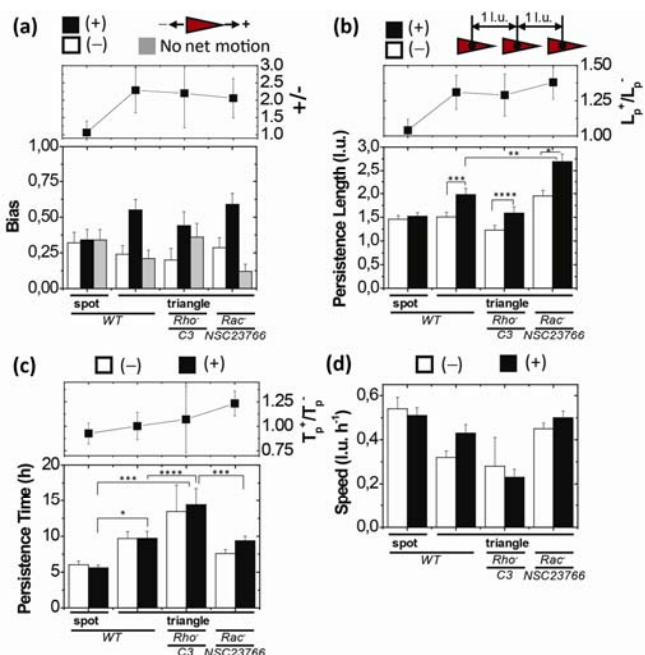

**Fig. 4.** Biophysical parameters studied as signatures of the directed NIH3T3 cell migration. **(a)** *(Lower)* Ratchet bias after 48 h. Three different behaviors were observed: net ratchet to the +,– directions and fluctuating NIH3T3 cells with no resulting net (null) motion. *(Upper)* +/− ratio. **(b, c, and d)** $L_p$, $T_p$ and v representation, respectively. In b, $L_p$ is larger for Rac$^-$, but a +, - shift in $L_p$ is observed for each condition (except for the control 'spot'), confirming that cells travel longer distances towards the + direction. (*1 l.u.* 'lattice unit' = *126.5 μm*). The upper panels in b and c represent the ratios between the + and − values of $L_p$ and $T_p$, for each condition, respectively. In d, the NIH3T3 cells move faster in the ratchet for Rac$^-$, whereas Rho$^-$ cells move slower than WT cells. (Data set: $(n;N)_{spot-WT}$=90;3. $(n;N)_{triangle-WT}$=121;5. $(n;N)_{C3}$=50;4. $(n;N)_{NSC23766}$=106;3). The data are presented as the mean ± SEM. *P<0.001; **P<0.005; ***P<0.01; ****P<0.05. (Student's *t*-test).

To test whether the fluctuation dynamics of protrusions for short timescales is sufficient to predict the persistence and asymmetry of cell trajectories on long timescale, we developed a mesoscopic model. The cell trajectories on homogeneous substrates have been shown in the literature to be well captured by persistent random walk models (33, 34). Such models differ essentially in their description of the ability of the cell to maintain its directionality, which is usually modeled as a memory kernel in the dynamics. We adapted similar ideas for our experimental conditions, in which the cell trajectories are discretized in lattice units (l.u.), with each elementary step being defined as a transition from one adhesive motif to a neighboring motif in either the + or – direction (in l.u.). In this discrete framework, we showed that a simple 1 step memory kernel was sufficient to capture the large-scale properties of the cell trajectories.

To do so, we introduce the conditional transition probabilities $\pi_{ji}$, where $i,j=+,-$ which are defined as the probability that a cell performs a step in the direction $j$, knowing that the previous step was performed in direction $i$ (see Fig. 5(a)). These quantities (only two of them $\pi_{++}$ and $\pi_{--}$ are independent due to the normalization conditions $\pi_{+i} + \pi_{-i} = 1$) describe the two effects responsible for the direction of migration: the asymmetry of the adhesive motifs, which as shown above, affects the protrusion activity, and the memory of the direction of the previous move, which is associated with cell polarity. The observation that $\pi_{i+} \neq \pi_{i-}$ (see Fig. 5(a, b)) clearly shows that the memory of the previous move influences the direction of the next move so that at least a 1 step memory is needed to model the cell trajectories. We now show that the main characteristics of the trajectories can be expressed in terms of $\pi_{ji}$ only. The analysis of the stationary state shows that the probability $\Pi_+$ of observing a step in the + direction (with no knowledge of the previous step) is given by

$$\Pi_+ = \frac{1-\pi_{--}}{2-(\pi_{++}+\pi_{--})} \quad (2)$$

The corresponding quantity $\Pi_-$ is then obtained by substituting $+\leftrightarrow-$ (see Supplemental Methods). The bias (measured experimentally as the +/− ratio) is then conveniently quantified as

$$\frac{\Pi_+}{\Pi_-} = \frac{1-\pi_{--}}{1-\pi_{++}} \quad (3)$$

In turn, the persistence length in the +/− direction is readily obtained using the model in the case of infinitely long trajectories as

$$L_p^\pm = \frac{1}{1-\pi_{\pm\pm}} \quad (4)$$

In experiments, the trajectories are finite, either because of the finite observation time or because of the cells pausing. These effects are accounted for by defining the probability q that a trajectory ends at each step. The parameter q can then be determined experimentally from the mean duration (in number of steps) of a trajectory, which is equal to 1/q. We then finally obtain the following expression for the persistence length, which is used to analyze the experimental data:

$$L_p^\pm = \frac{1}{1-(1-q)\pi_{\pm\pm}} \quad (5)$$

Figure 5(c) shows that using the measured values of $\pi_{ji}$ (see Fig. 5(b)), the predictions for both the +/− ratio (quantified by $\Pi_+/\Pi_-$) and the persistence lengths are in very good agreement with the observations. Note that only two independent measurements were needed to predict the long-term motion. For each condition, we measured the transition probabilities $\pi_{++}$ and derived the long-term parameters $L_p$ and +/−, which matched the experimental results remarkably well without adjustable parameters. Altogether, these results prove that cell trajectories are well described by a biased persistent random walk model with 1 step memory.

We next tested whether we could link the $\pi_{ji}$ probabilities to the protrusion activity and whether the step direction correlates quantitatively to the sign of the direction index $I_{dir}$ defined in Eq. 1. We used experimentally accessible parameters: $v_+$ (resp. $v_-$) and the corresponding probability that a protrusion is stabilized and eventually mediates a force denoted by $s_+$ (resp. $s_-$). Following the result that direction of motion depends on the direction of the previous move, we assume that quantities $s_i$, $v_i$, and $z_i$ also depend on the direction of the previous move and should be written as $s_{ji}$, $v_{ji}$, and $z_{ji}$, respectively (with $i,j=+,-$). We can then define the probability $p(n_{ji})$ that $n_{ji}$ efficient protrusions are generated at the j edge at each discrete step, which is given by the following Poisson distribution

$$p(n_{ji}) = \frac{(z_{ji}\tau_0)^{n_{ji}}}{n_{ji}!} e^{-z_{ji}\tau_0}$$

(6)

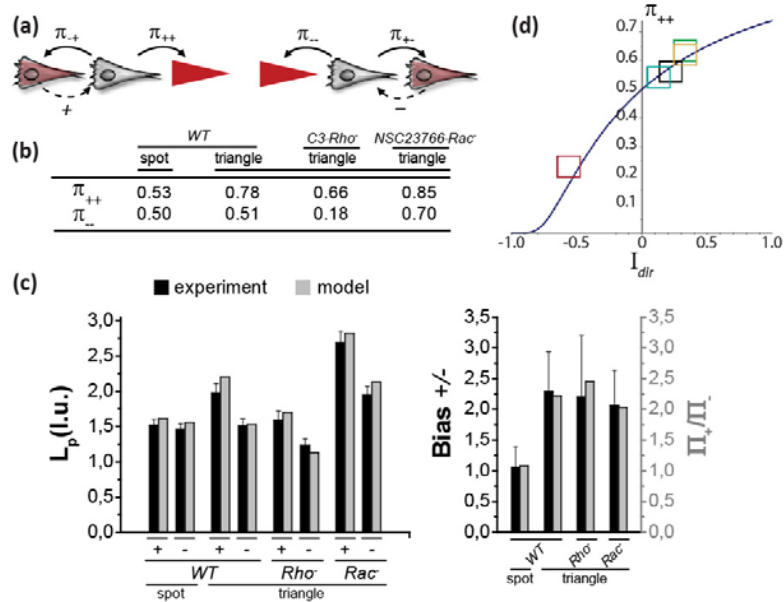

**Fig. 5.** Comparison between the model and experiments. **(a)** Schematic for $\pi_{ij}$; the cell first moves along the dashed lines and then along the solid lines. **(b)** Measurements for $\pi_{ij}$. **(c)** Comparisons between the model and experiments for the $L_p$ (in *l.u.*) and bias. **(d)** Predictions of $\pi_{++}$ as a function of the $I_{dir}$ given by Eq. 4. The quantity $(z_{ji} + z_{-ji})\tau_0$ is used as a fitting parameter. The q values used are as follows: WT: q=0.29; Rac⁻: q=0.24; Rho⁻: q=0.37; and spot: q=0.28. The experimental values are shown for each condition. *Key:* Red, rectangle-triangle-rectangle; Blue, rectangle-spot-rectangle; Black, *NSC23766*-ratchet; Yellow, *C3*-ratchet; and Green, *WT*-ratchet.

where $\tau_0$ is the duration that a cell stays on one motif, which we take to be independent of the direction of motion for the sake of simplicity. We then introduce an asymmetry in the efficient protrusions, $\Delta = n_{+i} - n_{-i}$. This quantity is the difference between random variables with Poisson distributions; it is distributed according to the so-called Skellam distribution (35):

$$P(\Delta) = e^{-(z_{+i}+z_{-i})\tau_0} \left(\frac{z_{+i}}{z_{-i}}\right)^{\Delta/2} I_\Delta\left(2\tau_0\sqrt{z_{+i}z_{-i}}\right) \qquad (7)$$

where $I_D(x)$ is the modified Bessel function of the first kind. Note that the average of $\Delta$ is, up to a normalization constant, the direction index $I_{dir}$ derived in Eq. 1. We then hypothesized that the direction of motion at each step is dictated by the edge with the largest number of stabilized protrusions. This condition, $\Delta \geq 1$, enables the following explicit calculation of $\pi_{ji}$:

$$\pi_{ji} = e^{-(z_{ji}+z_{-ji})\tau_0} \sum_{\Delta \geq 1} \left(\frac{z_{ji}}{z_{-ji}}\right)^{\Delta/2} I_\Delta\left(2\tau_0\sqrt{z_{ji}z_{-ji}}\right). \qquad (8)$$

Interestingly, this demonstrates that the $\pi_{ji}$ values, which fully determine the large-scale properties of the cell trajectories, as shown above, depend only on the mean number of stabilized protrusions $z_{ji}$ in each direction (and the timescale $t_0$). We finally note that the $\pi_{ji}$ values can be conveniently expressed in terms of $I_{dir}$ and an additional variable (for example, $(z_{ji} + z_{-ji})\tau_0$). The analytical prediction of Eq. 8 then shows that $\pi_{ji}$, and therefore the magnitude of the bias, critically depends on the sign of $I_{dir}$ (see Fig. 5(d)), confirming the experimental results (see Fig. 2(b)). To further test the model, we used the independent measures of $I_{dir}$ and the corresponding transition probabilities of Fig. 2(b) and 2(d) (which give the $\pi_{ji}$ values), which allowed for a direct comparison with the prediction of Eq. 8.

Figure 5(d) reveals the very good agreement between this theory and the experimental values in every condition. It is important to note that we used a *single* fitting parameter, $(z_{ji} + z_{-ji})\tau_0$. Altogether, these results validate our approach.

**Conclusions**
Although the molecular pathways that govern directional migration are fairly well understood, the role of stochasticity in protrusion activity has been overlooked so far. Our results show that the fluctuations of protrusions are key players in the physicochemical mechanism of directed NIH3T3 cell migration. We demonstrate that in our set-up, biased migration is based on the asymmetry in the protrusion activity, which is quantified by a simple index $I_{dir}$ that integrates the probabilities of protruding with the probabilities of stabilizing protrusions. From a cell biology perspective, our study suggests that cell polarity is not the only determinant of direction. Cell motion over short and long timescales has fluctuations, which play an important role. This does not mean that polarity is not important in setting direction but that cells integrate different factors when they probe their environment via protrusions. For example, a larger probability of finding an adhesive zone increases the probability of moving along this direction and potentially reversing the polarity. Careful measurements of the *z* parameters are then needed to estimate the probability for a cell to move in a given direction.

With our model, we could predict the long-term ratchet efficiency and persistence length using only two parameters as inputs. This highlights their important role in the physicochemical mechanism of directed cell migration. We anticipate that this simple framework will be useful for future studies on cell motility *in vitro* and *in vivo*.

**Acknowledgments**
The authors thank M. Labouesse (IGBMC, Strasbourg, France), F. Nédélec (EMBL, Heidelberg, Germany), M. Piel (Institut Curie, Paris, France), A. Bershadsky (Weizmann Institute, Rehovot, Israel) and all the members of the Riveline lab for stimulating discussions. A. Bornert is thanked for technical help. This work was supported by funds from the CNRS, the University of Strasbourg and the ci-FRC of Strasbourg.

# SUPPORTING MATERIAL

## Protrusion fluctuations direct cell motion

*David Caballero [1,2], Raphaël Voituriez [3,4], Daniel Riveline [1,2] ***

[1] Laboratory of Cell Physics, ISIS/IGBMC, Université de Strasbourg and CNRS (UMR 7006), 8 allée Gaspard Monge, 67083 Strasbourg, France

[2] Development and Stem Cells Program, IGBMC, CNRS (UMR 7104), INSERM (U964), Université de Strasbourg, 1 rue Laurent Fries, BP10142, 67400 Illkirch, France

[3] Laboratoire de Physique Théorique de la Matière Condensée, CNRS UMR 7600, Université Pierre et Marie Curie, 4 Place Jussieu, 75005 Paris, France

[4] Laboratoire Jean Perrin, CNRS FRE 3231, Université Pierre et Marie Curie, 4 Place Jussieu, 75005 Paris, France

*Corresponding author: Daniel Riveline; e-mail: riveline@unistra.fr, Laboratory of Cell Physics, ISIS/IGBMC, Université de Strasbourg and CNRS (UMR 7006), 8 allée Gaspard Monge, 67083 Strasbourg, France. TEL: +33 (0) 3 68 85 51 64; FAX: +33 (0) 3 68 85 52 32.

1. **Supplementary Material and Methods**

*Fibronectin concentration:* The fibronectin concentration (10 µg ml$^{-1}$) was selected in the following manner. The velocities of cells were measured for various concentrations on homogeneously coated microcontact-printed surfaces (see Fig. S3). As reported by Palecek *et al*, a dumbbell shape was recorded for the velocity-concentration plot with saltatory NIH3T3 cells in the lower region and crawling NIH3T3 cells in the higher density region (see ref. (23) in the main text). To obtain long displacements for long-term motion, we selected the higher speeds while keeping the cells in the crawling regime.

*Data analysis:* We measured biophysical parameters in the following manner. Cells trajectories were tracked manually by clicking on the centroid of each cell (Manual Tracking Plug-in, ImageJ) at 5 minute intervals for 48 hours. Points were connected to generate the migration path. Specifically, we used this method to extract $L_p$, $T_p$, v and $T_{pa}$ (a definition of these parameters is included in the Materials and Methods section in the main manuscript). $Nt^{-1}$ was recorded as the number of turns that cells performed during their full motion. Note that N is given by $1/(T_p+T_{pa})$; v is the ratio of $<L_p^i/T_p^i>$, where i=1..n cells, and v is different from $<L_p>/<T_p>$ because the distributions are not Gaussian. To classify the motions as +/−/null (no net motion), the cell positions at the start and the end of the movies were compared. 'No net motion' corresponded to cells returning to the original pattern location, although the cells had been fluctuating to the right and left during the acquisition. The data are provided as the mean ± SEM. Statistical analysis was performed using Student's *t*-test, and significance was accepted at P<0.05.

*Cell fixation and staining:* Cells were fixed with 3% paraformaldehyde (Sigma-Aldrich, France) at 37°C for 17 min or with methanol at -20°C for 10 min for centrosome fixation. Then, 0.5% Triton (Sigma-Aldrich, France) was added for 3 min to permeabilize the cells, and the samples were washed twice for 5 min with 1X PBS. For staining, we used Alexa Fluor 488-phalloidin (Molecular Probes) for actin, rabbit anti-pericentrin (Covance, UK) for the centrosome, and DAPI (4.6-diamidino-2-phenylidole, Sigma-Aldrich, France) for the nucleus. Focal contacts were stained with mouse anti-vinculin or anti-paxillin (Sigma-Aldrich, France), and microtubules were stained with mouse monoclonal anti-alpha tubulin (Sigma-Aldrich, France). The following secondary antibodies were used: goat anti-mouse conjugated with Cy3, Alexa 488 goat anti-rabbit (Molecular Probes, France) and Alexa Fluor 647 goat anti-mouse (Fisher Scientific SAS, France). The incubations with antibodies were performed for 45 min at room temperature in PBS. To stain the focal contacts on the pattern, incubation was conducted in PBS with 3% BSA. The stained samples were observed with an inverted fluorescence microscope (Eclipse Ti, Nikon, Japan) combined with a Photometrics CoolSNAP HQ$^2$ camera. We used a pre-centered fiber illuminator (C-HGFIE, Nikon, Japan) and an X60 oil objective (Nikon, Japan) with a numerical aperture of 1.3.

**2. Supplemental Figures and Legends.**

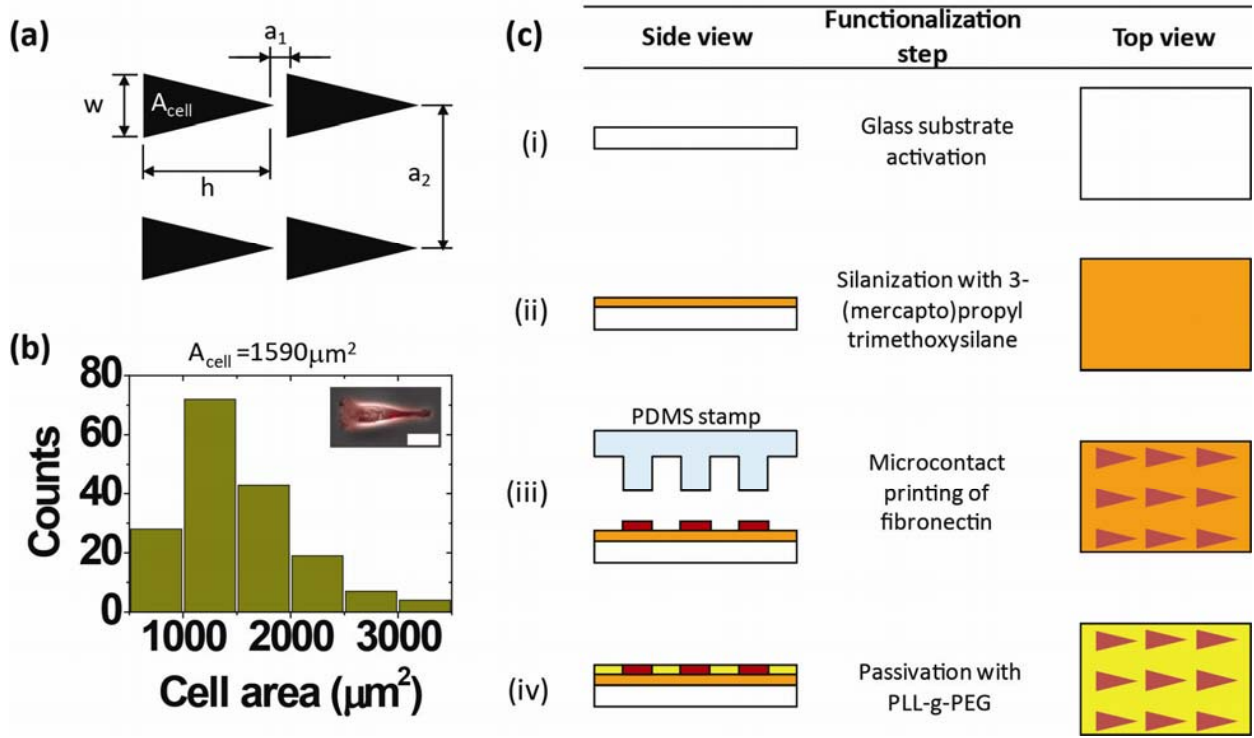

**Fig. S1. Ratchet design and protocol for its fabrication. (a)** Micropattern design showing the different values used. $w=30$ $\mu m$; $h=106$ $\mu m$; $a_1=20.5$ $\mu m$; $a_2=200$ $\mu m$ and $A=1590$ $\mu m^2$. **(b)** Optimization of the micropattern area. Cells were deposited on a microcontact printed glass coverslip with a 10 µg ml$^{-1}$ fibronectin concentration and were allowed to spread. The area of individuals cells (n=176 cells) was measured, and the averaged $A_{cell}=1590\pm54\mu m^2$ was selected for the motif design. The data are shown as the mean ± SEM. Inset, a cell completely spread on a single asymmetric fibronectin motif. Scale bar = 35 µm. **(c)** Functionalization protocol (see Methods). *(i)* Glass coverslips, #1, were thoroughly cleaned and chemically activated with *'Piranha'* solution. *(ii)* Glass coverslips were silanized by vapor phase deposition with 3-(mercapto)propyl trimethoxysilane and placed in an oven at 65°C for 4 hours. *(iii)* The microcontact printing technique was used to deposit fibronectin patterns on the substrates. *(iv)* Non-functionalized regions were passivated with poly(L-lysine)-g-poly(ethylene glycol) (PLL-g-PEG). Samples were stored in PBS (pH 7.4) at 4°C for at least 30 minutes prior to use.

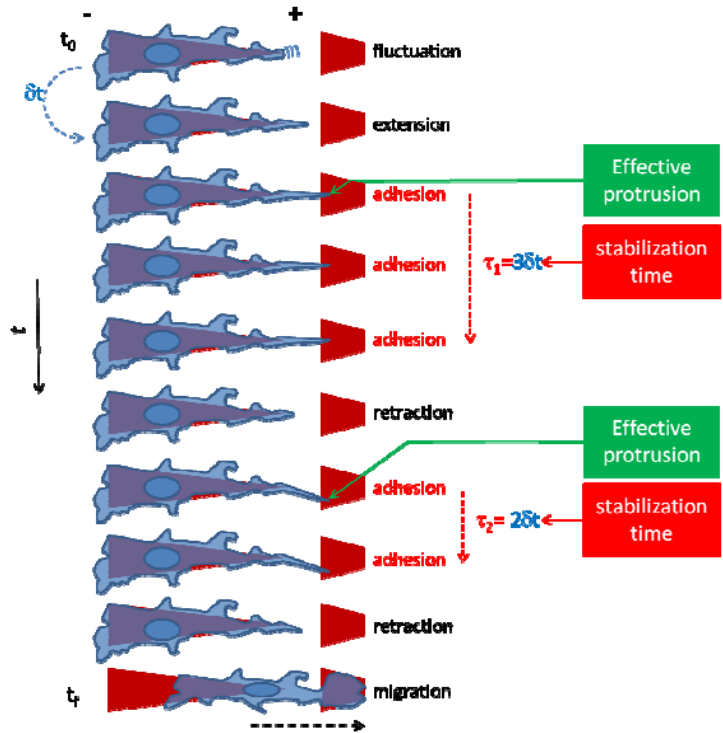

**Fig. S2. Scheme showing the definitions of ν and τ.** A cell protrusion fluctuates, grows and adheres (effective protrusion) on the neighboring motif during $\tau_1$ and $\tau_2$ before retracting and migrating towards the + direction. δt represents the time between acquisitions, and $t_0$ and $t_f$ represent the start and end of the experiment, respectively.

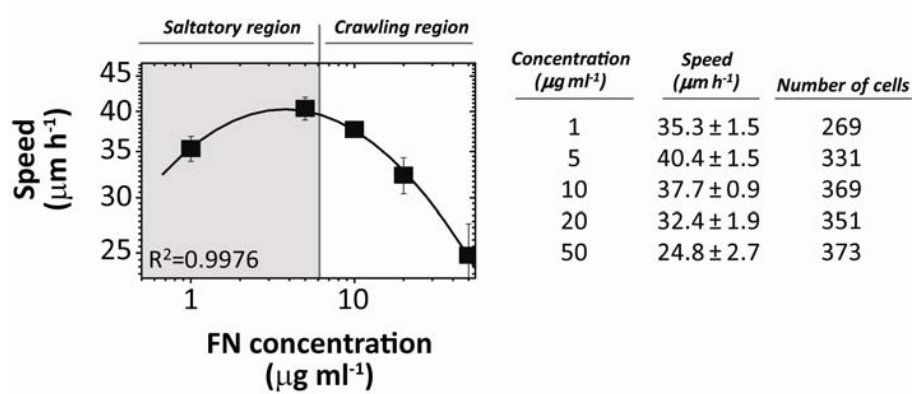

**Fig. S3. Representation of NIH3T3 cell migration speed as a function of fibronectin coating concentration.** *(Left)* Variation of the mean speed of NIH3T3 cells on surfaces that were homogeneously coated with fibronectin (FN). The speed is measured by tracking the centroid of each individual cell: we compute the initial and final points where NIH3T3 cells are moving straight. We observed two different regimes: the first one ranged from *1≤[FN]<6 μg ml$^{-1}$* (grey zone) with saltatory cells, and the second one ranged from *6≤[FN]≤50 μg ml$^{-1}$* (white zone) with crawling cells. The data correspond to N=6 independent experiments for at least 269 NIH3T3 cells per condition for a total of n=1693 cells. *(Right)* Table showing the speed obtained for each fibronectin concentration with the number of cells analyzed per condition. The values are shown as the mean ± SEM.

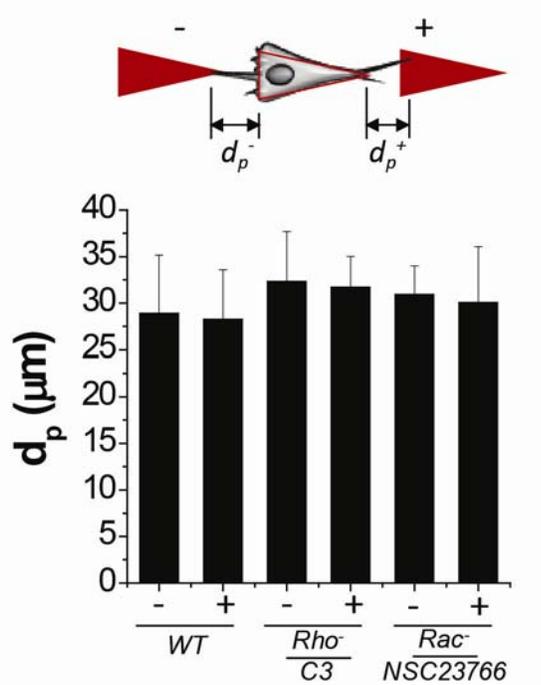

**Fig. S4. Protrusion length extension.** Average protruding distance $d_p$ in the + and − directions. NIH3T3 cells extend filopodia protrusions of similar lengths in all directions for each condition. (Data set: *WT*=885; *C3-Rho$^-$* = 1084 and *NSC23766-Rac$^-$* = 1084 analyzed filopodia. N=3). The data are presented as the mean ± SD.

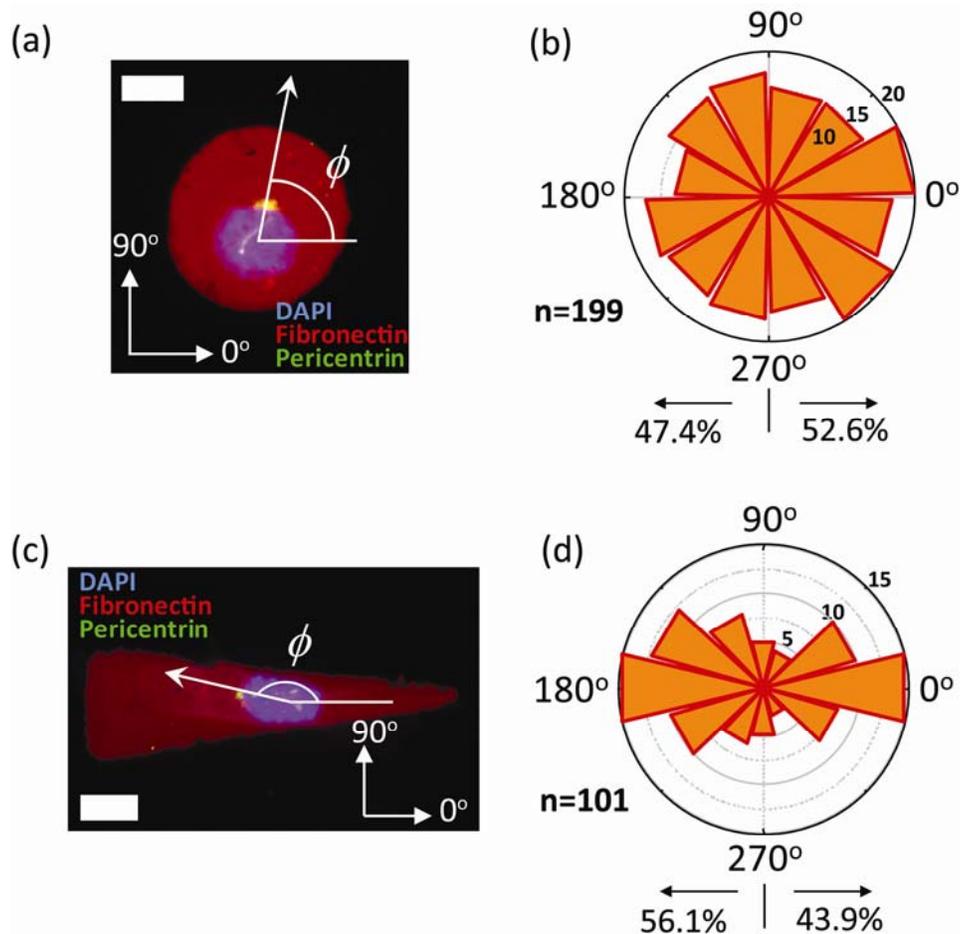

**Fig. S5. Internal polarity caused by the asymmetric fibronectin motif. (a)** NIH3T3 cells were allowed to spread prior to fixation. They were stained for the nucleus (blue) and centrosome (green). The centrosome positions were measured with respect to the nucleus (see Methods). Scale bar: 15 μm. **(b)** Circular histogram of the angular distributions for the centrosome with respect to the nucleus centroid (n=199 cells) for cells plated on fibronectin spots; the distribution is isotropic. **(c)** Fixation and staining for the nucleus (blue) and centrosome (green) of NIH3T3 cells on rhodamine-labelled asymmetric fibronectin (10 μg ml$^{-1}$) triangles. The centrosome positions were measured with respect to nucleus. Scale bar: 15 μm. **(d)** Circular histogram of the angular distributions of the centrosome with respect to the nucleus centroid. (n=101 cells). The histogram shows a distribution of centrosomes along the longer cell axis and a bias towards the wide edge.

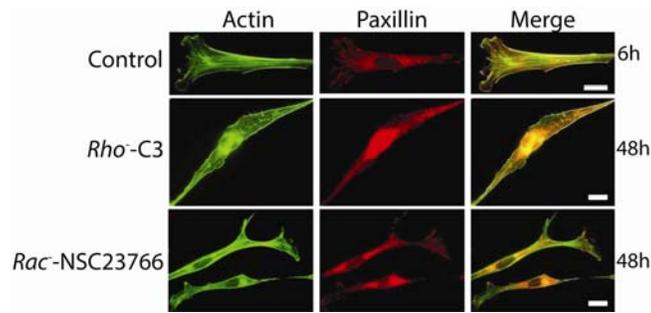

**Fig. S6. Inhibiting the Rho and Rac pathways clearly affect the cytoskeleton.** NIH3T3 cells were deposited on a microcontact printed glass coverslip with a 10 µg ml$^{-1}$ fibronectin concentration. We compared the effects of inhibiting the Rho and Rac pathways on the cell cytoskeleton via treatments with the inhibitors C3 transferase (80 nM) and NSC23766 (100 µM), respectively. NIH3T3 cells were labeled for actin and paxillin. The right column shows the incubation time at fixation; note that similar phenotypes were observed for each condition throughout the experiments. Scale bar = 10 µm.

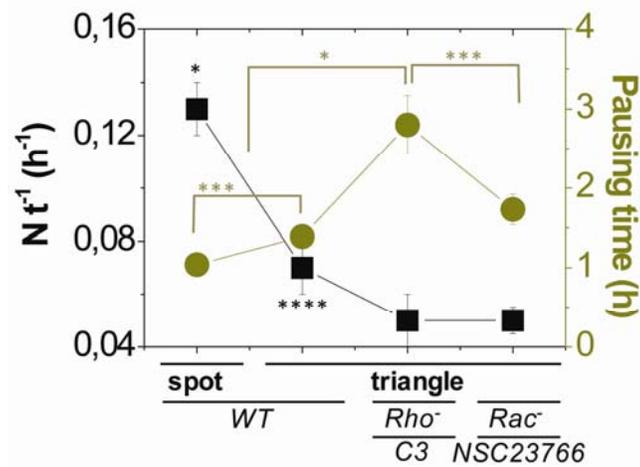

**Fig. S7. Characterization of the long-term cell motion.** *(Left)* Number of changes in the direction per unit time for all conditions. The cells switch direction more often for the *WT* spot condition. The data suggest a directional motion that lasts longer for the ratchet configuration with Rho[-] and Rac[-]. *(Right)* Pausing time ($T_{pa}$) prior to a change in direction for all conditions. For the Rho[-] condition, the cells show the highest $T_{pa}$ value. The data are presented as the mean±SEM. *$P<0.001$; ***$P<0.01$; ****$P<0.05$. (Student's *t*-test).

## 3. Supplementary Model Description

### 3.1 Probabilistic model for the direction index $I_{dir}$ - first step motion

We present a simple model that predicts the probability $p_+$ that a protrusion is stabilized in the + direction rather than the − direction for a cell on a 1d lattice of triangular adhesive patterns. We show that direction index $I_{dir}$ quantifies the asymmetry in this quantity: $I_{dir} = p_+ - p_-$. The width and height of a triangular region are $h$ and $w$, respectively (we follow the notations in Fig. S1). The gap distance between the two nearest adhesive regions is $a_1$. Cells extend protrusions to the nearest adhesive regions on the right and on the left. The probability distribution that an extended protrusion has the length $l$ has the form

$$\psi(l) = \lambda e^{-\lambda l} \tag{1}$$

where $\lambda^{-1}$ is the average length of the protrusions. This exponential form for the probability distribution was shown experimentally by Xia *et al* (FASEB J. **22**:1649-1659 (2008)). Cells extend $n_+$ (multiple) protrusions towards the right and $n_-$ protrusions towards the left at the same time (where $n_+ \neq n_-$ for the case of cells on triangular patterns because of cell polarization) with frequency $1/\tau_p$. The probability $\psi_+$ that a protrusion that extended towards the right touches the adhesive region at the nearest neighbor is derived by integrating $\psi(l)$ in the adhesive regions at the nearest neighbor and has the form

$$\psi_+ = g_+ e^{-\lambda a_1}, \tag{2}$$

where $g_+$ is the geometrical factor, which depends on h and w and which we do not aim at determining here. Similarly, the probability that a protrusion that extended towards the left touches the adhesive region at the nearest neighbor has the form

$$\psi_- = g_- e^{-\lambda a_1}. \tag{4}$$

The inequality $g_+ \neq g_-$ reflects the fact that the accessible adhesive area $A_\pm$ is different for the protrusions that extended towards the + and − directions. Because of the relatively large gap between adhesive regions, the cells must exert forces that are larger than a threshold value to move from one adhesive region to another. This requires coordinated dynamics between the protrusions and retraction and therefore requires that the protrusions are stabilized. Here, we assume that this stabilization occurs with a Poisson process of rate $\beta$. Our experiments show that the stabilization time, during which a protrusion is stabilized on adhesive regions, depends on whether this protrusion is extended towards the right or left, most likely because of cell polarization. A protrusion that touches the (nearest) adhesive regions at the right is stabilized with probability

$$s_+ = 1 - e^{-\beta \tau_+}, \tag{5}$$

where $\tau_+$ is the stabilization time for protrusions extended towards the right direction. On the left, the probability $s_-$ has a stabilization time $\tau_-$. With this model, the total number of protrusions that are extended towards the right is $n_+ \tau_0 / \tau_p$, where $\tau_0$ is the duration that a cell stays in the same adhesive region before moving to the next region. For these protrusions, the ratio of protrusions that touch the nearest adhesive region is $\psi_+$, and those protrusions that touch the adhesive region are stabilized and become efficient with probability $s_+$. The probability that a protrusion is stabilized at the right rather than the left thus has the form

$$p_+ = C v_+ \tau_0 s_+, \tag{6}$$

where $v_+ = n_+ \psi_+ / \tau_p$ is the frequency of protrusions that touch the nearest adhesive region.

Similarly, the probability that a protrusion is stabilized at the left rather than the right is derived in the form

$$p_- = Cv_-\tau_0 s_-, \tag{7}$$

where the normalization constant $C$ is deduced from

$$C^{-1} = v_+ s_+ \tau_0 + v_- s_- \tau_0, \tag{8}$$

which implies that a protrusion is stabilized either on the + or − side. One then obtains the following:

$$p_+ - p_- = \frac{v_+ s_+ - v_- s_-}{v_+ s_+ + v_- s_-}$$

$$\approx \frac{v_+ \tau_+ - v_- \tau_-}{v_+ \tau_+ + v_- \tau_-} = I_{dir} \tag{9}$$

The last equation on the right hand side was derived by assuming that the rate $\beta$ is much smaller than the inverse of the stabilization time (for both protrusions towards the right and left). This equation is indeed equal to direction index $I_{dir}$ defined in Eq. 1 of the main text with $z_+ = v_+ s_+$ and $z_- = v_- s_-$. This provides a probabilistic interpretation of the direction index.

### 3.2 Persistent random walk model – long-term motion

The model aims at connecting the observed persistence and asymmetry in the +/− directions of trajectories to the fluctuation dynamics of protrusions. The persistent random walk model is defined as follows. A cell trajectory is discretized, with each elementary step defined as a transition from one adhesive motif to a neighboring motif, in either the + or − direction. We introduce the conditional transition probabilities $\pi_{ji}$, where $i,j = +,-$; these probabilities are defined as the probability that a cell performs a step in the direction $j$, knowing that the previous step was performed in direction $i$. Normalization then imposes $\pi_{+i} + \pi_{-i} = 1$; thus, only two quantities (for example, $\pi_{++}$ and $\pi_{--}$) are independent. The $\pi_{ji}$ *describe* the two effects responsible for the direction of migration: the asymmetry of the adhesive motifs and the direction of the previous move. The observation that $\pi_{i+} \neq \pi_{i-}$ (see Figure 5 of the main text) clearly shows that the memory of previous move, which is likely to affect the internal organization of organelles, influences the direction of the upcoming move. We now show that the main characteristics of the trajectories can be expressed in terms of the $\pi_{ji}$. The probability $\Pi_+(t)$ of observing a step in the + direction at time step t (with no knowledge of the previous step) obeys the following Markov dynamics

$$\Pi_+(t+1) = \pi_{++}\Pi_+(t) + \pi_{+-}\Pi_-(t) \tag{10}$$

The corresponding dynamics for $\Pi_{-+}(t)$ is then obtained by substituting $+ \leftrightarrow -$. The analysis of the stationary state defined by $\Pi_+(t+1) = \Pi_+(t) = \Pi_+$ yields

$$\Pi_+ = \frac{1 - \pi_{--}}{2 - (\pi_{++} + \pi_{--})} \tag{11}$$

The corresponding quantity $\Pi_-$ is then obtained by substituting $+ \leftrightarrow -$.

4. **Supplementary Movie Titles and Legends.**

**Movie S1 – Protrusion probing activity on an FN ratchet.** Movie of an individual NIH3T3 fibroblast showing the different steps of protrusion (filopodia) activity prior to cell migration. The cell starts to fluctuate after spreading and polarizing on an adhesive fibronectin ratchet. This generates a cell front and a cell rear. The colored arrows (blue and green) highlight the probing of the protrusions on the neighboring adhesive motifs. The yellow arrows indicate the start of the cell spreading and migration and the retraction at the back of the cell. Acquisition time: 1 image/30 seconds. Time in hh:mm:ss. (Scale bar: 50 μm).

**Movie S2 – Protrusion probing activity on a triangle with 2 symmetric rectangles.** Protrusion (filopodia) activity of an NIH3T3 cell spread on an FN triangle with rectangle-shaped neighboring motifs on both sides. The colored arrows (blue and green) highlight the probing of the protrusions on the neighboring adhesive motifs. The yellow arrows indicate the start of cell spreading and migration. Acquisition time: 1 image/30 seconds. Time in hh:mm:ss. (Scale bar: 50 μm).

**Movie S3 – Protrusion probing activity on a spot with 2 symmetric neighboring rectangles.** Protrusion (filopodia) activity of an NIH3T3 cell spread on an FN spot with rectangle-shaped neighboring motifs on both sides. The colored arrows (blue and green) highlight the probing of the protrusions on the neighboring adhesive motifs. The yellow arrows indicate the start of cell spreading and migration. Acquisition time: 1 image/30 seconds. Time in hh:mm:ss. (Scale bar: 50 μm).

**Movie S4 – Fluctuating NIH3T3 cell on a fibronectin spot pattern.** Movie of an individual NIH3T3 fibroblast fluctuating for 48 hours from one fibronectin spot (in red) to the neighboring ones, crossing the PLL-g-PEG blocked regions, and changing its direction of migration. Acquisition time: 1 image/5 minutes. Time in hh:mm. (Scale bar: 100 μm).

**Movie S5 – NIH3T3 cell motion on a sequence of triangular fibronectin patches.** Movie of an individual NIH3T3 fibroblast for 30 hours plated on a sequence of micropatterned fibronectin triangular patches (in red). The NIH3T3 cell fluctuates right and left at the beginning of the movie from one fibronectin triangle to the neighboring ones, crossing the PLL-g-PEG blocked regions. Then, the cell migrates directionally in the + direction. Acquisition time: 1 image/5 minutes. Time in hh:mm. (Scale bar: 100 μm).